\documentclass[twocolumn,prb]{revtex4-2}
\usepackage{graphicx,amssymb,soul,color,amsmath,bm}
\usepackage{epstopdf,hyperref}
\usepackage{braket,dsfont,nicefrac}
\usepackage[mathcal]{euscript}
\usepackage[normalem]{ulem}
\usepackage[utf8]{inputenc}

\hypersetup{colorlinks=true,citecolor=blue,linkcolor=magenta}

\sethlcolor{yellow}
\allowdisplaybreaks

\begin{document}

\title{Ferromagnetism and doublon localization in a Wannier-Hubbard chain}

\author{Samuel Milner}
\affiliation{Department of Physics, Northeastern University, Boston, Massachusetts 02115, USA}
\author{Phillip Weinberg}
\affiliation{Department of Physics, Northeastern University, Boston, Massachusetts 02115, USA}
\author{Adrian Feiguin}
\affiliation{Department of Physics, Northeastern University, Boston, Massachusetts 02115, USA}

\begin{abstract}
We derive a ``Wannier-Hubbard'' model consisting of an array of overlapping atomic orbitals interacting via a local Coulomb interaction. Transforming to an orthogonal Wannier basis set, the resulting Hamiltonian displays long range hopping and interactions, with new terms such as correlated hopping and ferromagnetic direct exchange, among others. We numerically study the one-dimensional version of the model at half-filling using the density matrix renormalization group (DMRG) method, unveiling a rich phase diagram as a function of the interaction $U$ and the overlap $s$ with metallic, and ferromagnetic phases, separated by a ferrimagnetic region. Our results indicate a path toward understanding new emergent phases under pressure and beyond standard model Hamiltonians.     
\end{abstract}

\maketitle

\section{Introduction}
Despite enormous progress in understanding the chemistry and complex many-body phenomena taking place in correlated materials, predicting magnetic order, particularly in metals, remains a challenging endeavor. 
Paradoxically, the origin of ferromagnetism in strongly correlated materials is more puzzling than antiferromagnetism. In fact, in insulators, magnetic interactions are almost universally antiferromagnetic and, in the context of the much studied Hubbard model\cite{Hubbard1963,tasaki1997,EsslerBook}, antiferromagnetic interactions arise naturally from super exchange \cite{Anderson1950, Anderson1959, Anderson1963}. However, in the cuprate La$_4$Ba$_2$Cu$_2$O$_{10}$ (La422) ferromagnetic direct exchange leads to an unexpected transition to a FM phase below 5K\cite{Mizuno1990, Masuda1991, PIEPER1994,Ku2002}. This kind of exceptions cannot be accounted for by the Nagaoka mechanism \cite{Nagaoka1966} nor flat band Hamiltonians \cite{Tasaki1998}, pointing toward the incompleteness of the Hubbard model when it comes to explaining the emergence of ferromagnetism in correlated insulators. This issue was already raised a while back in Refs.\onlinecite{Strack1994,Amadon1996}, where it was observed that a ferromagnetic exchange would arise if one considered nearest neighbor Coulomb terms. 
In particular, these models assume a Wannier basis and do not explicitly account for the effects of the overlap between atomic orbitals on the direct exchange. The full magnetic exchange consists of direct exchange and superexchange contributions. While the first one is ferromagnetic, the sign of the second one is determined by the Goodenough-Kanamori rules\cite{Goodenough1955}. As a rule of thumb, the exchange interaction for the case of overlapping non-orthogonal orbitals is antiferromagnetic in order to reduce the kinetic energy, while non-overlapping (orthogonal) orbitals lead to a ferromagnetic interaction in order to minimize the potential energy (paradigmatic examples of both cases are the Heitler-London model for the hydrogen molecule\cite{Heitler1927} and Hund's rule, respectively \cite{MattisBook}). 

\begin{figure}
    \centering
    \includegraphics[width=\linewidth]{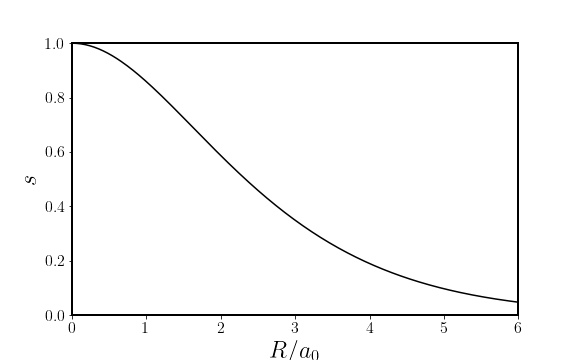}
    \caption{Overlap between two $1s$ orbitals in a hydrogen dimer as a function of the inter-atomic distance, in units of the Bohr radius.}
    \label{fig:overlap}
\end{figure}

The effects of the overlap on the matrix elements can be substantial, as we shall see, significantly affecting the physics of the problem. As a simple illustration, the overlap between two $1s$ orbitals in a hydrogen dimer are plotted in Fig.\ref{fig:overlap} as a function of the inter-atomic distance $R$ in units of the Bohr radius. The actual value for the dimer is $R=2.4$, implying an overlap close to $s=0.5$. As a matter of fact, this value is so large that, as we shall see, the formalism we present in this work fails. In this case one needs to resort to a multi-orbital treatment including, for instance, $2s$ states in the same spirit as configuration interaction calculations in quantum chemistry to account for the orbital fluctuations\cite{Hirsch1993H2}.  

\begin{figure}
    \centering
    \includegraphics[width=\linewidth]{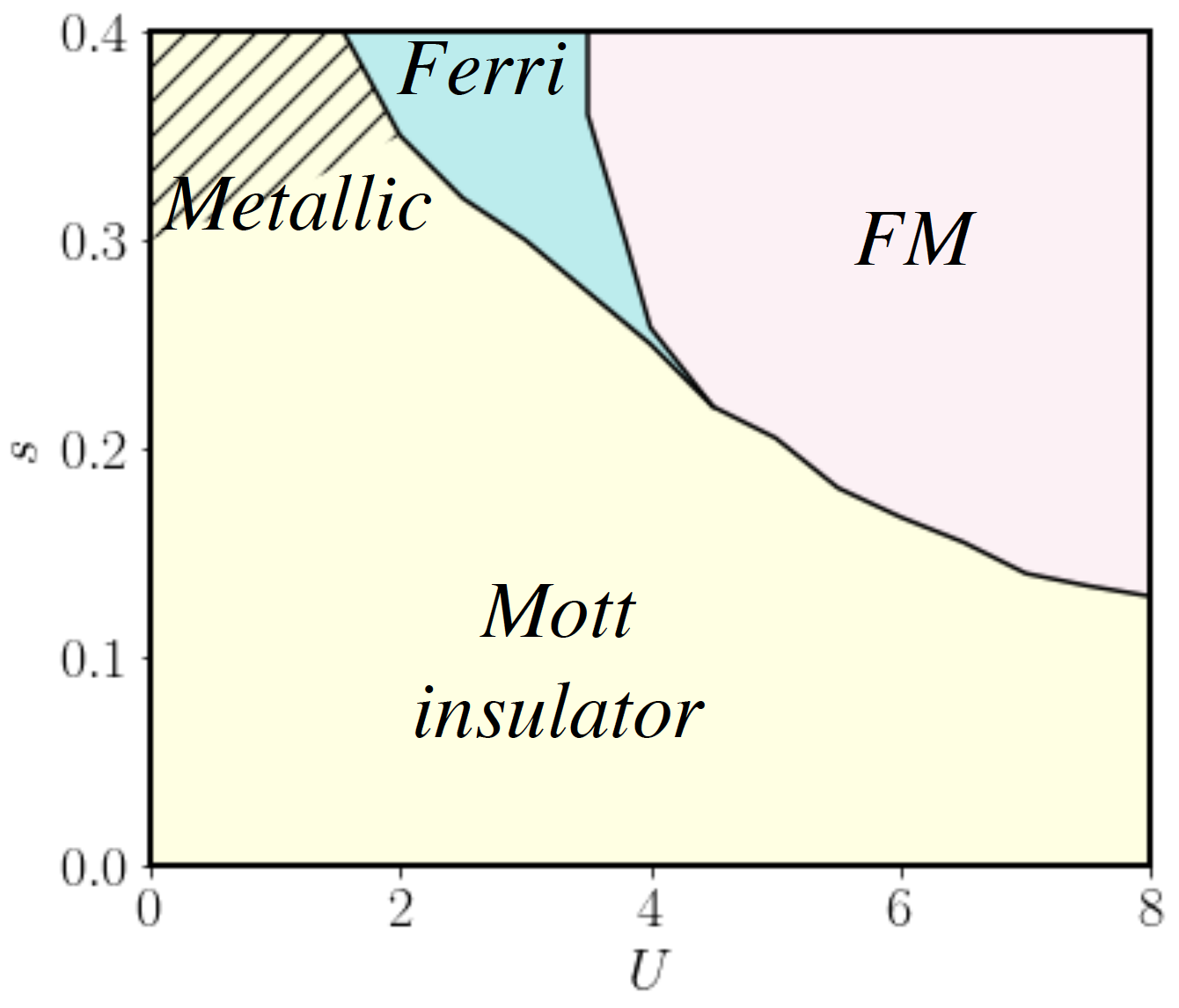}
    \caption{Phase diagram of the Wigner-Hubbard chain for chain with $L=32$ sites as a function of the overlap $s$ and the interaction $U$, obtained with DMRG by locating level crossings in the spectrum and measuring the total spin $S$. We find a Mott insulating phase, a ferromagnetic (FM) phase with spin $S=L/2$, a metallic region for small $U$ and larger $s$, and an intermediate ferrimagnetic region. }
    \label{fig:phase_diagram}
\end{figure}

In this work we revisit the Hubbard model with new light, by including the terms that naturally emerge from the orthogonalization to a Wannier basis \cite{Anderson1959,Anderson1963,Irkhin1998}. These terms include not only second neighbor hopping, but also direct exchange, pair hopping and correlated hopping. Besides the obvious loss of particle-hole symmetry, the latter contribution leads to a dramatic increase of the mass of the doublons and very narrow upper Hubbard bands, as we show in our results. In addition, a natural competition between ferromagnetic exchange and antiferromagnetic super exchange gives rise to an insulating ferromagnetic phase in the ground state phase diagram (Fig.~\ref{fig:phase_diagram}). We demonstrate this behavior by carrying out exact diagonalization and density matrix renormalization group (DMRG) \cite{White1992,White1993,Schollwock2005,Schollwock2011,Feiguin2013a} calculations in one-dimensional chains.

The manuscript is organized as follows: In sec. \ref{sec:hubbard} we derive the Wannier-Hubbard Hamiltonian by starting from a basis of non-orthogonal orbitals; in sec.\ref{sec:results} we numerically study the phase diagram and characterize the different phases, including gaps, correlations, and spactral functions. Finally, we close with a summary and conclusions.

\section{Wannier-Hubbard model}\label{sec:hubbard}
\subsection{From atomic to Wannier orbitals}
In our derivation of the Wannier-Hubbard model we start from the assumption that matrix elements beyond those acting locally on a site/atom are very small and can be ignored; in second quantization the Hamiltonian for a single band reads as a conventional Fermi-Hubbard model:
\begin{equation}
	H = -\sum_{ij,\sigma} t_{ij} \tilde{c}^\dagger_{i\sigma}\tilde{c}_{j\sigma} +U\sum_i \tilde{c}^\dagger_{i\uparrow}\tilde{c}_{i\uparrow}\tilde{c}^\dagger_{i\downarrow}\tilde{c}_{i\downarrow},
\end{equation}
where the constants $t_{ij}$ represent the hopping matrix elements and the parameter $U$ is the on-site Coulomb integral. Since we consider only one atomic orbital per site and one is typically interested in the insulating regime, we ignore long range terms that would also include the exchange integral. In this expression, the $\tilde{c}$ orbitals represent a non-orthogonal basis of atomic orbitals (for instance, $1s$ orbitals in a hydrogen chain\cite{Fermann1994, Hirsch2013, Hirsch2014, Motta2017, Motta2020, Stoudenmire2017, Hirsch2021, Sawaya2022}. For this reason, they do not satisfy canonical anticommutation rules:
\begin{equation}
\left\lbrace \tilde{c}_i^\dagger,\tilde{c}_j\right\rbrace  = S_{ij},
\end{equation}
where $S$ is the overlap matrix $S_{ij}=\langle\emptyset|\tilde{c}_i\tilde{c}^\dagger_j|\emptyset\rangle$ (we notice that when adding a spin index, the different spin sectors are decoupled and they can be treated independently). In order to transform to a new canonical representation obeying Fermi statistics, we carry out a \emph{non-unitary} transformation (see Appendix \ref{app:canonical} and Ref.\onlinecite{jorgensen1981second}):
\begin{equation}
	\tilde{c}_i = \sum_\nu (S^{1/2})_{\nu i} c_{\nu}.
\end{equation}

\begin{figure}
    \centering
    \includegraphics[width=\linewidth]{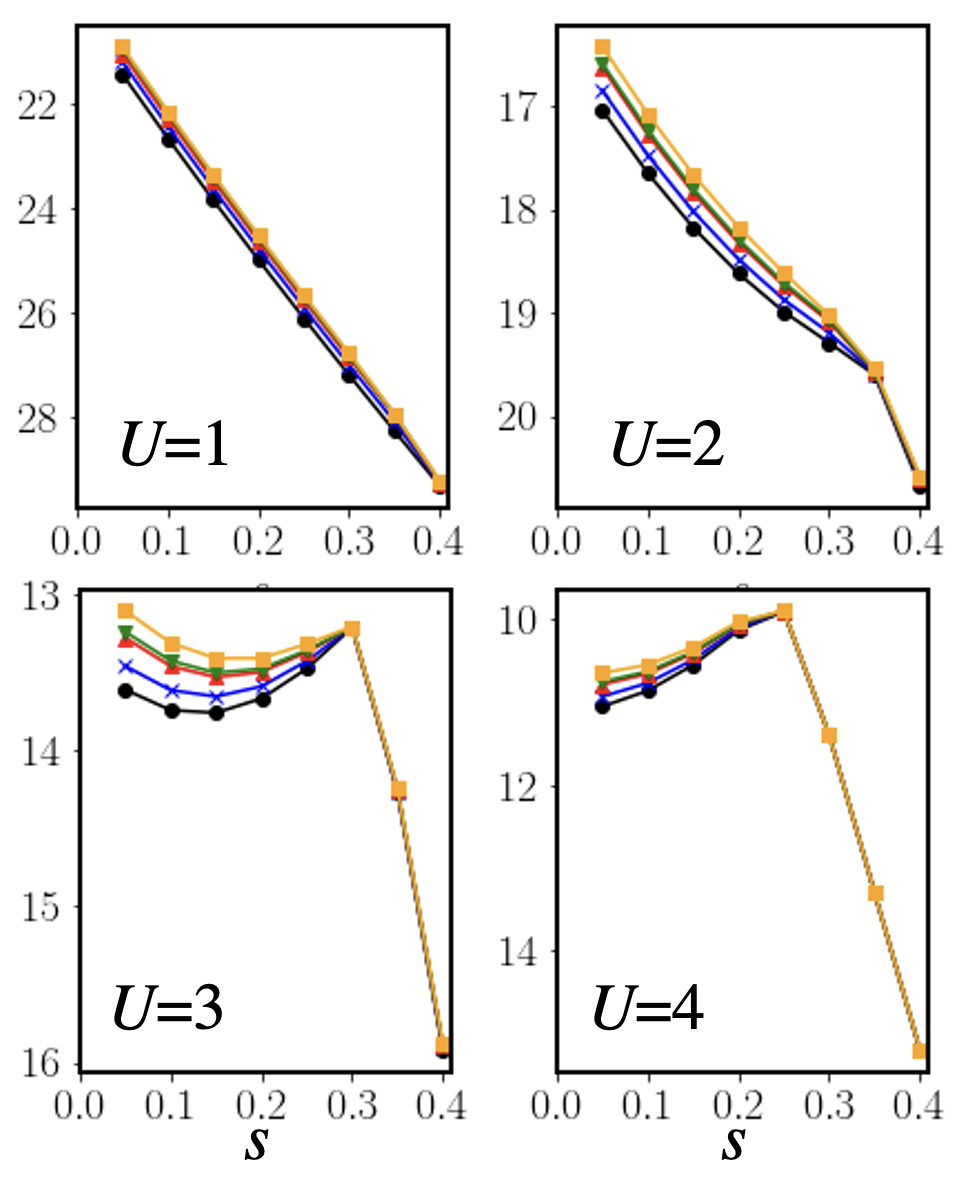}
    \caption{Lowest five energy eigenvalues for a chain with $L=20$ as a function of $s$ and for several values of $U$. The degeneracies indicate a multiplet ground state, signature of either ferri or ferromagnetism. }
    \label{fig:levels}
\end{figure}

The Hamiltonian in terms of the orthonormal basis defined by the canonical Fock operators reads:
\begin{equation}
	H =  \sum_{\mu\nu,\sigma}\tilde{t}_{\mu\nu} c^\dagger_{\mu\sigma}c_{\nu\sigma} + U\sum_{\mu\nu\gamma\delta} u_{\mu\nu\gamma\delta} c^\dagger_{\mu\uparrow}c_{\nu\uparrow}c^\dagger_{\gamma\downarrow}c_{\delta\downarrow}.
\end{equation}
where we have parametrized the two-body terms with a $4$-tensor 
\begin{equation}
u_{\mu\nu\gamma\delta}=\sum_{i}(S^{1/2})_{\mu i}(S^{1/2})_{\nu i}(S^{1/2})_{\gamma i}(S^{1/2})_{\delta i}
\end{equation}
and introduced a new hopping matrix $\tilde{t}_{\mu\nu}  = (S^{1/2} t S^{1/2})_{\nu\mu}$. Notice that despite its apparent simplicity, the interaction now contains $N^4$ terms, where $N$ is the system size. 

\begin{figure}
    \centering
    \includegraphics[width=\linewidth]{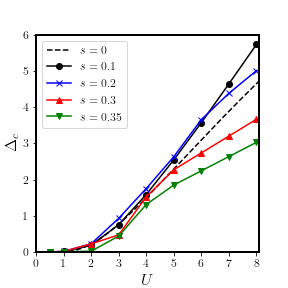}
    \caption{Charge gap as a function of $U$ for several values of the overlap $s$. We also include the exact Bethe Ansatz results for $s=0$ as a comparison. }
    \label{fig:gap}
\end{figure}

\subsection{The Wannier-Hubbard chain}
We now consider a chain geometry in which atomic orbitals overlap only with their nearest neighbors with $\langle\emptyset| \tilde{c}_i\tilde{c}^\dagger_{i\pm1}|\emptyset\rangle = s$, such that $(S)_{i,i\pm 1} = s$ and $(S)_{i,i} = 1$. For a periodic chain with translational invariance, the matrix $S$ can be diagonalized using a Fourier transform giving eigenvalues: $\sigma_k = 1+2 s \cos(k)$. This simple solution in terms of plane waves allows one to compute $S^{1/2}$ exactly:
\begin{equation}
	\left(S^{1/2}\right)_{\mu j} = \frac{1}{L}\sum_k e^{i(j-\mu)k}\sqrt{1+2 s \cos(k)}.
\end{equation}
The new kinetic energy term in the canonical basis will be given as:
\begin{equation}
    H_0 = -\sum_{\mu,\sigma,\delta=1,2} \tilde{t}(\delta) \left(c^\dagger_{\mu\sigma} c_{\mu+\delta,\sigma} + \mathrm{h.c.}\right),
    \label{eq:h0}
\end{equation}
corresponding to a tight-binding chain with nearest neighbor hopping $\tilde{t}({\delta=1})=t$ and second neighbor hopping $\tilde{t}({\delta=2})=s$ (there is a constant shift in the chemical potential that we have ignored). This long range hopping automatically destroys particle-hole symmetry\cite{Fabrizio1996, Arita1998, Daul1998, Daul2000, japaridze2007a, Tocchio2010, Mishmash2015}.

\begin{figure}
    \centering
    \includegraphics[width=\linewidth]{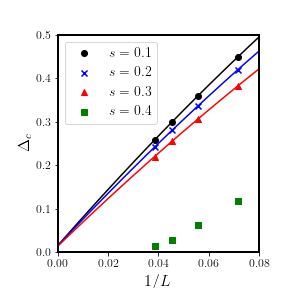}
    \caption{Finite size scaling of the charge gap for $U=1$ and different values of $s$. Symbols are DMRG data and lines are extrapolations to the thermodynamic limit using a quadratic polynomial in $1/L$.}
    \label{fig:gap_vs_l}
\end{figure}

One can now carry out an expansion of the interactions terms in powers of $s$ (see Appendix \ref{app:expansion}). The largest contribution remains the local on-site density-density interaction that occurs when $\mu=\nu=\delta=\gamma=0$ which is $\mathcal{O}(1)$. In the next order $\mathcal{O}(s)$ we find correlated hopping terms obtained by plugging $\mu=\nu=\delta=0$ and $\gamma=1$:
\begin{equation}
	H_1=U\frac{s}{2}\sum_{\mu\sigma\delta=\pm 1}\left( n_{\mu\sigma}c^\dagger_{\mu\bar{\sigma}}c_{\mu+\delta\bar{\sigma}}+\mathrm{h.c.} \right)
	\label{eq:h1}
\end{equation}
At order $\mathcal{O}(s^2)$ we get a density-density interaction, a pair hopping term and a direct exchange term:
\begin{eqnarray}
	H_2 &=& U\frac{s^2}{2}\sum_\mu \left[\frac{1}{2}n_{\mu}n_{\mu+1} - 2\vec{S}_\mu \cdot\vec{S}_{\mu+1} \right. \nonumber \\ 
	&+& \left.\left(c^\dagger_{\mu\uparrow}c^\dagger_{\mu\downarrow}c_{\mu+1\downarrow}c_{\mu+1\uparrow} + \mathrm{h.c.}\right) \right]
	\label{eq:h2}
\end{eqnarray}
In addition, there are other terms of order $\mathcal{O}(s^2)$ that involve operators on more than two sites:
\begin{eqnarray}
	H_3 &=& U\frac{s^2}{4}\sum_\mu \left(-c^\dagger_{\mu\uparrow}S^-_{\mu+1}c_{\mu+2\downarrow}-
	c^\dagger_{\mu\downarrow}S^+_{\mu+1}c^\dagger_{\mu+2\uparrow}+ \right. \nonumber \\
	&+& c^\dagger_{\mu\uparrow}n_{\mu+1\downarrow}c_{\mu+2\uparrow}+
	c^\dagger_{\mu\downarrow}n_{\mu+1\uparrow}c_{\mu+2\downarrow}+ \\
	&+& \left. c^\dagger_{\mu\uparrow}c_{\mu+1\downarrow}c_{\mu+1,\uparrow}c^\dagger_{\mu+2\downarrow}+
	c_{\mu\uparrow}c^\dagger_{\mu+1\downarrow}c^\dagger_{\mu+1,\uparrow}c_{\mu+2\downarrow}+\mathrm{h.c.} \right) \nonumber
	\label{eq:h3}
\end{eqnarray}
Longer range terms of similar form will enter with progressively higher powers of the overlap $s$. 
At this point we notice in Eq.(\ref{eq:h2}) that the spin-spin exchange interaction has a negative sign, indicating that it is ferromagnetic. This is the dominant direct exchange contribution because we have completely ignored the exchange integral between neighbors and only included the local Coulomb interaction in the Hubbard Hamiltonian. This, of course, it a quite artificial situation: if the overlap is large, one would also expect the Coulomb integral between neighboring sites to be significant, and not zero\cite{Kishore1971, SPATEK1977}. This model describes a situation in which the direct exchange originates purely from the overlap between orbitals. However, one should also recognize the presence of an antiferromagnetic super exchange $J_{AF}\sim 4t^2/U$. As we find below, a competition between these two contributions will ultimately determine the spin order in the ground state.

\section{Numerical Results}
\label{sec:results}
\subsection{Phase diagram and Mott insulator transition}
In order to understand the implication of the new terms of the Hamiltonian and how the physics differs from that of the conventional Hubbard model, we carried out DMRG calculations on chains different lengths at, and away from half-filling. For validation, we compared our results in small systems against exact diagonalization using QuSpin\cite{QuSpin}. We initially considered the full Hamiltonian with $N^4$ terms, but in order to scale up our simulations to larger system sizes we used the truncated model to order $\mathcal{O}(s^2)$, keeping enough states to guarantee a truncation error below $10^{-6}$. 

Focusing first on the Mott insulating phase at half-filling ($N=L$) we explore the phase diagram as a function of $U$ and $s$, shown in Fig.\ref{fig:phase_diagram}. The boundaries between the different regions were obtained by comparing the ground state energy to the energies of the fully saturated state with $S^z_{Tot}=L/2$ as well as the first excited state with $S^z_{Tot}=1$. An $(L+1)$-fold degenerate ground state would indicate a ferromagnetic ground state; otherwise, a lower degeneracy would mean that the ground state is a multiplet with a smaller total spin, a ferrimagnet. A typical behavior is shown in Fig.\ref{fig:levels} where we plot the lowest five eigenvalues as a function of the overlap $s$ for several values of $U$ and $L=20$.  As mentioned above, we expect that the competition between direct exchange $J_{FM}=Us^2/2$ and indirect exchange $J_{AF}=4t^2/U$ will lead to ferromagnetism becoming energetically favorable for $s \gtrsim \sqrt{8}t/U$. In reality this simple expression, although a rule of thumb to serve as intuition, is not quite satisfied since it does not include the gains due to the changing bandwidth or potential energy in each phase. 

\begin{figure}
    \centering
    \includegraphics[width=\linewidth]{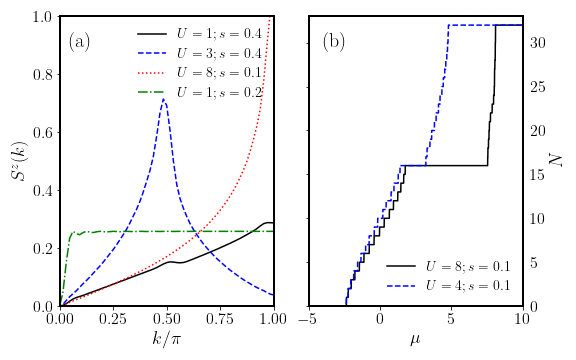}
    \caption{(a) Static longitudinal spin structure factor $S^z(q)$ for different values of $U$ and $s$ representative of the four phases found in the phase diagram. Results are for a chain with $L=32$ sites in the spin $S^z=0$ sector. (b) Particle number $N$ as a function of chemical potential $\mu$ demonstrating the particle hole asymmetry and the flattening of the upper Hubbard band for $L=16$, $s=0.1$ and $U=4, 8$. }
    \label{fig:mu}
\end{figure}

Interestingly, besides a ferromagnetic (FM) phase at large values of $U$ and $s$ with spin $S_{Tot}=L/2$, we also encounter an intermediate regime with ferrimagnetic order and total spin $S_{Tot}<L/2$. In addition, we explored the possibility of a gapless metallic regime by studying the behavior of the charge gap $\Delta_c=E_0(N+1)+E_0(N-1)-2E_0(N)$, where $E_0$ is the ground state energy for a given particle number. In Fig. \ref{fig:gap} we show results for this quantity extrapolated to the thermodynamic limit using a quadratic fit in $1/L$. In the ferromagnetic phase the gap grows linearly with $U$. At small values of the overlap $s$ the behavior of the charge gap is similar to the one of the conventional Hubbard chain, which can be calculated exactly using the Lieb-Wu Bethe Ansatz formula\cite{Lieb1968}. The gap is exponentially small for small $U$ and there is no metal-insulator transition at finite values of the interaction. For this reason, it is difficult to pinpoint the boundaries of the metallic phase for large $s$, if any. However, we observe a markedly different scaling behavior of the gap as a function of $1/L$ in the shaded range of parameters in the phase diagram, that suggests that the gap is indeed zero for small $U$, as seen in Fig.\ref{fig:gap_vs_l}. Interestingly, similar physics is observed in the Hubbard chain with second neighbor hopping $t_2$ \cite{Daul2000}. In both cases, the stabilization of a metallic phase can be attributed to the increased kinetic energy with larger $t_2$, or $s$ in our model and coincides with the development of two minima in the spectrum.

In addition, we calculated the longitudinal spin structure factor $S^z(q)$ defined as:
\[
S^z(q)=\frac{1}{L^2}\sum_{lm}e^{iq(l-m)}\langle S^z_lS^z_m \rangle.
\]
Resulls for values of $U$ and $s$ representative of the different phases are shown in Fig.\ref{fig:mu}(a). While we observe a trivial behavior in the ferromagnetic phase, we find a sharp peak at $2k_F=\pi$ in the Mott insulating ground state indicating dominant antiferromagnetic correlations, as also observed in the conventional Hubbard chain. The results in the metallic phase do not differ significantly from those on the non interacting $t_1-t_2$ tight-binding problem, except for the kink at $k=k_F=\pi/2$. As $U$ increases into the ferrimagnetic phase, the correlations develop a peak at $k=\pi/2$ indicating an instability toward incommensurate spin order with power law decay.     

We also investigated the possibility of other phases away from half-filling, namely, pairing and charge density wave (CDW). For this purpose, we calculated the particle number $N$ as a function of chemical potential $\mu$ for different values of $s$ and $U$. Pleateaus would indicate a charge gap, hence a possible CDW, while steps in $\Delta N=2$ would suggest the possibility of pairing. We do not identify such anomalous behaviors, ruling out those possibilities. However, we clearly observe the particle-hole anisotropy manifested as a shrinking of the bandwidth of the upper Hubbard band. In fact, the upper Hubbard band is almost flat for small $s \sim 0.1$, as shown in Fig.\ref{fig:mu}(b). 

\subsection{Photoemission spectrum}

In order to further characterize the different phases we also study the momentum resolved single particle spectrum of the model using the time-dependent DMRG method\cite{white2004a,daley2004,Feiguin2005,vietri,Paeckel2019}. 
The spectral function is defined as the imaginary part of the two time correlator:
\[
G(x,t)=-i\langle O^\dagger_{x}(t)O_{L/2}(t=0) \rangle.
\]
A Fourier transform from space and time to momentum and frequency yields the desired momentum-resolved Green's function.
In practice, we carry out two simulations: one for the operator $O=c_\uparrow$ that would produce the spectral function for the empty states below the Fermi energy, and a second one for $O=c^\dagger_\uparrow$ for the occupied states above $E_F$. At half-filling, these calculations would result into the lower and upper Hubbard bands, respectively. The presence of three-site terms in the Hamiltonian prevents us from using a Suzuki-Trotter decomposition of the evolution operator. Therefore, we use instead a time targeting approach with a Krylov expansion of $\exp{(-iHt)}$, as described in Ref.\onlinecite{Feiguin2005,vietri}. Same as for the ground state calculations, we keep enough DMRG states to ensure that the truncation error remains below $10^{-6}$.

\begin{figure}
    \centering
    \includegraphics[width=\linewidth]{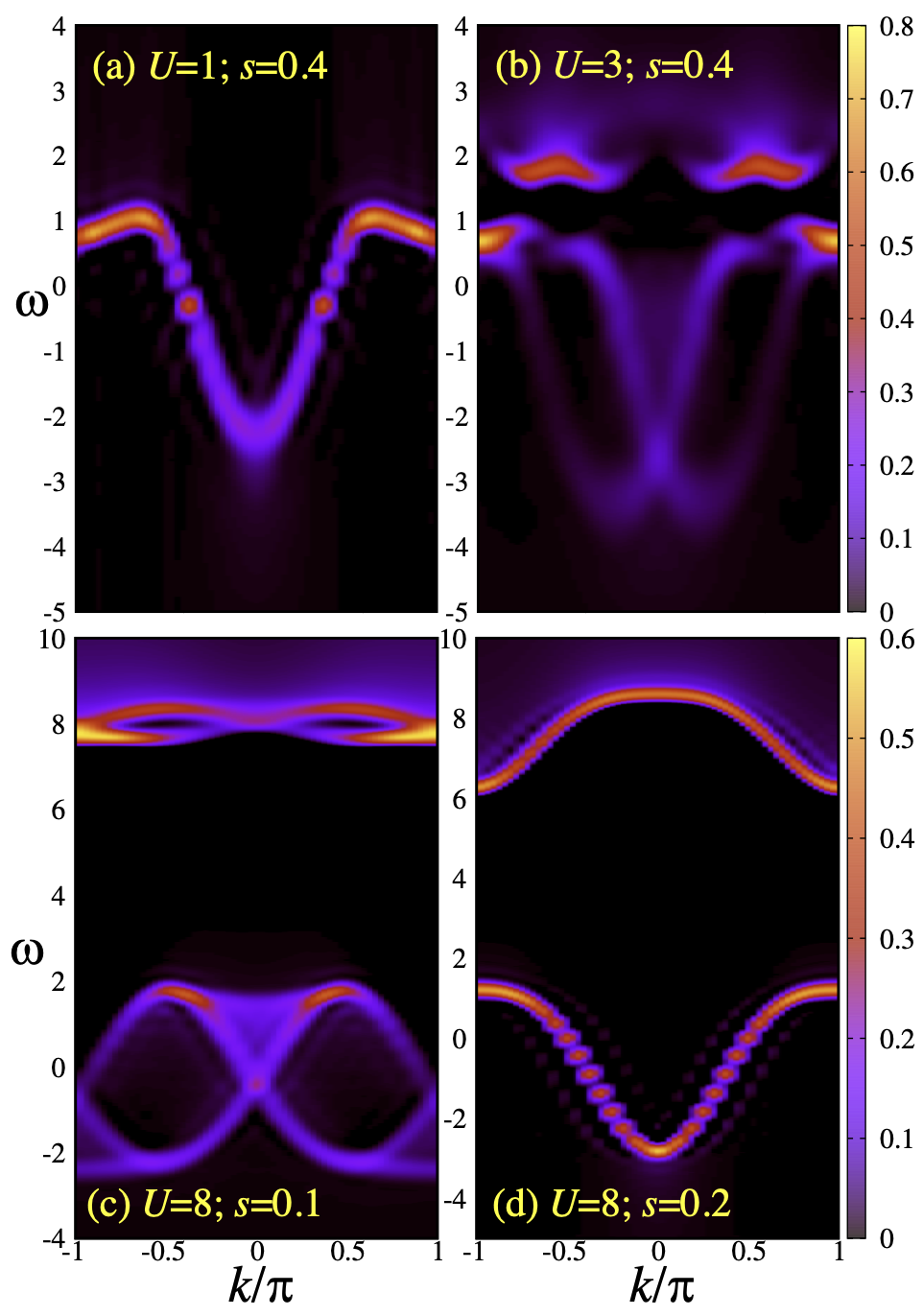}
    \caption{Spectral function for $L=32$ at half-filling for parameters in the (a) metallic; (b) ferrimagnetic; (c) Mott insulating; (d) ferromagnetic phases.}
    \label{fig:spectrum}
\end{figure}

Fig.\ref{fig:spectrum} shows results at half-filling for the same values of $U$ and $s$ as in Fig.\ref{fig:mu}(a). In panel (a) we recognize the gapless dispersion of the metallic state modified by effective second neighbor hopping due to $s$. In this case the Fermi energy is at $E_F=0$, while in the gapped cases it lies inside tha gap. In Fig.\ref{fig:spectrum}(b), the incommensurate features in the ferrimagnetic case suggest a broken translational symmetry, with the spinon dispersion shifting and centered at $k=\pi$ instead of $k=0$. Correlations (not shown) indicate a possible weak trimerization, with the excess spin sitting in between trimers. In the Mott insulating phase, Fig.\ref{fig:spectrum}(c), the lower Hubbard band is practically unaffected and displays spinon and holon features with a bandwidth $W\sim 4t$. However, the upper Hubbard band, corresponding to doublon excitations is very narrow. We attribute this effect to the correlated hopping term in Eq.(\ref{eq:h1}), that penalizes the motion of doublons and can be significant, due to the prefactor $Us/2 = 0.4$ for these parameters. This would give doublons an effective hopping $t_d \sim 0.6$ and a corresponding bandwidth $W \sim 4t_d = 2.4$. This is larger than the measures $W\sim 1$ but our simple intuitive arguments obviously ignore other possible effects that arise at higher order in $s$. Finally, in the ferromagnetic phase, Fig.\ref{fig:spectrum}(d), holes move freely in a ferromagnetic background and the band is well described by a non-interacting dispersion $\epsilon_h(k)=-2t\cos{k}-2s\cos{2k}$. Similarly, doublons can be interpreted as up-spin electrons moving in a fully polarized background with the opposite orientation, with a dispersion of the form $\epsilon_d(k)=2(t-Us/4)\cos{k}+2(s-Us^2/2)\cos{2k}$, where the extra contributions originate from Eqs.(\ref{eq:h1}) and (\ref{eq:h3}).

\section{Conclusions}

To summarize this work, we derived a ``Wannier-Hubbard'' Hamiltonian describing a lattice of non-orthogonal atomic orbitals, where electrons interact only locally via a Coulomb repulsion, akin to the conventional Hubbard model. As a consequence of the overlap between adjacent orbitals, the interactions in the Wannier basis become effectively long range. 
The resulting model contains the same type of terms that would emerge from explicitly including all the Coulomb integrals between neighboring sites if one worked directly in a Wannier basis, as originally done by Hubbard \cite{Hubbard1963,Strack1994}. Typically, those terms are considered independent and varied as free parameters\cite{Aligia1999}. In our derivation, we start from a basis of atomic orbitals by only considering the dominant local Coulomb integral $U$. The overlap between adjacent orbitals propagates the interactions between orthogonal Wannier orbitals and the additional contributions naturally arise. In our formulation, all the matrix elements can be directly associated to $U$ and the overlap $s$, quantities that can be obtained from relatively straightforward integrals and derived from first principles. This provides a more natural intuition and can be used to understand, for instance, the effects of pressure in terms of the spatial dependence of $s$, as illustrated by Fig.\ref{fig:overlap}.   

In particular, we find that the most important contributions originate from a second-neighbor hopping, correlated hopping and a ferromagnetic spin exchange.  
The competition between super exchange and direct exchange drives the system to a ferromagnetic phase for large values of the interaction $U$ and the overlap $s$, also observed in Ref.\onlinecite{Amadon1996}. We also encounter a metallic regime for small $U$ and large $s$ and an intermediate phase with ferrimagnetic order.
The ferromagnetic interaction in our model arises in the orthogonal basis due to the finite overlap between neighboring atomic orbitals. 
Its net effect, together with the Coulomb repulsion in Eq.(\ref{eq:h2}), is to push the particles apart: electrons prefer to arrange in a ferromagnetic state in which they occupy localized orthogonal orbitals and are completely disentangled, hence minimizing the potential energy. Notice that this is in effect the equivalent to Hund's rule at a macroscopic scale. Since particles are completely localized, the charge gap in this state grows linearly with $U$. 

In addition, correlated hopping for doublons induces a flattening of the upper Hubbard band. This appears as a sharp step in the $N$ versus $\mu$ curve. When this occurs, instead of phase segregating the system undergoes a transition to a ferromagnetic state. Interestingly, ferromagnetism has also been observed in a generalized Hubbard model with correlated hopping and without exchange\cite{Gagliano1995}, indicating that this term is important in driving the transition.   

Our model illustrates the rich physics that can be realized by simply considering additional but fundamental effects such as the overlap between adjacent atomic orbitals, that are important in materials under pressure, as experimentally demonstrated\cite{Mizuno1990, Masuda1991, PIEPER1994}. Our results indicate a path toward understanding the emergence of ferromagnetism in Mott insulators and potentially other novel phases beyond those encountered in simple model Hamiltonians such as the conventional Hubbard model. 

\acknowledgements
The authors acknowledge the U.S. Department of Energy, Office of Basic Energy Sciences for support under grants No. DE-SC0014407 (AEF and SM) No. DE-SC0019275 (PW). We are grateful to A. Chernyshev for illuminating discussions.

\appendix

\section{Non-canonical Fock operators}\label{app:canonical}

In this section we briefly review the orthogonalization procedure used to obtain a new canonical basis set, following Ref. \onlinecite{jorgensen1981second}.
Consider a set of non-orthogonal orbitals $|\phi_i\rangle=\tilde{c}^\dagger_i|\emptyset\rangle$ that are localized a discrete set of lattice sites labeled by $i$, such as atomic orbitals. We denote the overlap matrix of these states with $S$ such that
\begin{equation}
\langle\emptyset|\tilde{c}_i\tilde{c}^\dagger_j|\emptyset\rangle = S_{ij}.
\end{equation}
Slater-determinants using these non-orthgonal orbitsl have an equivalent representation in terms of Fock operators that obey the following non-canonical anti-commutation relations:
\begin{equation}
	\left\lbrace\tilde{c}_i^\dagger,\tilde{c}_j\right\rbrace  = S_{ij}
\end{equation}
To construct the canonical Fock operators can use a linear transformation of the non-orthogonal Fock operators:
\begin{equation}
	c_\nu = \sum_{i} V_{\nu i}\tilde{c}_i.
\end{equation}
We require that these operators satisfy canonical Fermi statistics:
\begin{equation}
	\left\lbrace c_\mu^\dagger,c_\nu\right\rbrace = \sum_{ij} V^*_{\mu i}S_{ij} V_{\nu j} = \delta_{\mu\nu}
\end{equation}
Since $S$ is symmetric and positive definite we can chose $V=S^{-1/2}$ to satisfy the last equality. The inverse transformation from the orthogonal basis to  non-orthogonal operators is given by:
\begin{equation}
	\tilde{c}_i = \sum_\nu (S^{1/2})_{\nu i} c_{\nu}.
\end{equation}
We notice that when adding a spin index, the different spin sectors are decoupled at the level of the $S$ and they can be treated independently. 

\section{Expansion in $s$}\label{app:expansion}

In the limit $L\rightarrow\infty$ we can write the sum over momentum as an integral. Given the translational invariance of the problem we find that the matrix elements do not depend on the absolute values of $j$ and $\mu$ but only the distance between the two indices $j-\mu$. We will define this distance as $\rho$. The final result for the integral is given by:
\begin{equation}
	\left(S^{1/2}\right)_{j+\rho,j}=\int_{-\pi}^\pi \frac{dk}{2\pi} e^{-ik\rho}\sqrt{1+2s\cos(k)}
\end{equation}

Our goal will be to evaluate this expression to leading order in $s$ as a function of $\rho$. First we assume that $s<1/2$ such that we can Taylor expand the square root inside the integral:
\begin{equation}
	\sqrt{1+2s\cos(k)} = \sum_{r=0}^\infty \binom{1/2}{r} (2s)^r\cos^r(k)
\end{equation}
Plugging this expression into the original integral and swapping the order of the sum and the integral we find:
\begin{equation}
	\left(S^{1/2}\right)_{j+\rho,j}= \sum_{r=0}^\infty\binom{1/2}{r} (2s)^r \int_{-\pi}^\pi \frac{dk}{2\pi} e^{-ik\rho}\cos^r(k)
\end{equation}
Now we can evaluate the integral using properties of complex exponentials:
\begin{equation}
	\int_{-\pi}^\pi \frac{dk}{2\pi} e^{-ik\rho}\cos^r(k) = \frac{1}{2^r} \sum_{m=0}^r \binom{r}{m}\delta_{2m,r+\rho}
\end{equation}
Finally, by using the fact that $cos^r(k)$ is an even function one can show that $\left(S^{1/2}\right)_{j+\rho,j}=\left(S^{1/2}\right)_{j-\rho,j}$ hence,
\begin{equation}
	\left(S^{1/2}\right)_{j+\rho,j} = \sum_{r=0}^\infty\binom{1/2}{r} s^r\sum_{m=0}^r \binom{r}{m}\delta_{2m,r+|\rho|}
\end{equation}
Now to get the expansion to leading order, we notice that when $r<|\rho|$ the second sum over $m$ is always $0$, hence the leading order term in the expansion will always correspond to $r=|\rho|$, in which case the sum reduces to $1$. 
\begin{equation}
	\left(S^{1/2}\right)_{j+\rho,j} = \binom{1/2}{|\rho|} s^{|\rho|} + \mathcal{O}(s^{|\rho|+1})
\end{equation}

Using this relationship we can derive the terms in the Hamiltonian corresponding to each power of $s$. In particular, for the local Coulomb interaction on site $i$, we find:
\begin{widetext}
\begin{multline}
	\tilde{c}^\dagger_{i\uparrow}\tilde{c}_{i\uparrow}\tilde{c}^\dagger_{i\downarrow}\tilde{c}_{i\downarrow} = \sum_{\mu\nu\delta\gamma} \left[(S^{1/2})_{i+\mu, i}(S^{1/2})_{i+\nu, i}(S^{1/2})_{i+\gamma, i}(S^{1/2})_{i+\delta, i}\right] c^\dagger_{i+\mu\uparrow}c_{i+\nu\uparrow}c^\dagger_{i+\gamma\downarrow}c_{i+\delta\downarrow} \approx\\
	\sum_{\mu\nu\delta\gamma} \binom{1/2}{\mu}\binom{1/2}{\nu}\binom{1/2}{\delta}\binom{1/2}{\gamma}s^{\mu+\nu+\gamma+\delta} c^\dagger_{i+\mu\uparrow}c_{i+\nu\uparrow}c^\dagger_{i+\gamma\downarrow}c_{i+\delta\downarrow}
\end{multline}
\end{widetext}

\bibliography{references,1d-Hubbard-t1-t2,hydrogen}
\end{document}